# Internal Symmetry of Space-Time Connections with Torsion

David R. Bergman[a]
*Morristown*, NJ

**Abstract**

In this brief article an internal symmetry of a generic metric compatible space-time connection, metric and generalized volume element is introduced. The symmetry arises naturally by considering a space-time connection containing a generic torsion tensor, and would otherwise be missed in the absence of torsion. When the transformation is applied to the Hilbert-Einstein action it is shown that by a choice of gauge all possible field theories arising from the Hilbert-Einstein action are equivalent to the standard theory of gravity described by general relativity.

[a] Author's permanent contact info: davidrbergman@essc-llc.com, davidrbergman@yahoo.com

## 1. Introduction

In The Meaning of Relativity Fifth Edition [1] it is shown that the Riemann, Ricci and Einstein curvature tensors are invariant under a transformation of the connection,

$$\overline{\Gamma}^{\alpha}{}_{\lambda\beta} = \Gamma^{\alpha}{}_{\lambda\beta} + \delta^{\alpha}_{\beta} \partial_{\lambda} f(x^{\mu}) \tag{1}$$

In equation (1), $f$ is an arbitrary function of coordinates. The transformation is applied to the connection only, the metric tensor, coordinates and other structures defined on the manifold, $M$, remaining unchanged. The reader is referred to [1] for details regarding this transformation and the proof of invariance of the curvature tensor. In [1] this is referred to as a "lambda" transform, $\lambda$. The proof of the symmetry does not rely on the choice of connection and does not require any restrictions on the symmetry of the connection with respect to index exchange. Einstein used this fact to argue that, in general, the connection may contain an anti-symmetric component, torsion, that is not present in the Christoffel symbols. The torsion field was initially assumed to be due to an anti-symmetric component of the metric tensor.

A conformal transformation is a transformation applied to the metric tensor and the affine parameter used to describe auto-parallel or geodesic curves on $M$. For a complete description of this transformation the reader is referred to the text, General Relativity by Robert Wald [2]. The conformal transformation is given by,

$$g_{\mu\nu} \to \Omega^2 g_{\mu\nu} \tag{2a}$$

$$g^{\mu\nu} \to \Omega^{-2} g^{\mu\nu} \tag{2b}$$

$$\sqrt{\det(g_{\mu\nu})} \to \Omega^N \sqrt{\det(g_{\mu\nu})} \tag{2c}$$

where the notation of Wald [2] has been used and $\dim(M) = N$. The effect of the conformal transformation on the connection is given by,

$$\overline{\Gamma}^{\alpha}{}_{\lambda\beta} = \Gamma^{\alpha}{}_{\lambda\beta} - \delta^{\alpha}_{\lambda}\partial_{\beta}\ln\Omega - \delta^{\alpha}_{\beta}\partial_{\lambda}\ln\Omega + g_{\lambda\beta}g^{\alpha\rho}\partial_{\rho}\ln\Omega \qquad (3)$$

Note that no part of the torsion tensor is affected by this transformation. All terms in equation (3) arise from the metric induced portion of the connection, the Christoffel symbols.

In this work these two transformations are applied to the connection field and it is shown that when the torsion tensor is subjected to an independent gauge transformation the connection is completely invariant with respect to the complete set of transformations, referred to here as the "Omega" transform, $\Omega$. Once the connection is shown to be invariant the invariance of all curvature tensors is almost trivial, the exception being the scalar curvature since that involves a multiplication of the Ricci tensor by the contravariant metric tensor. Next, the behavior of the Hilbert-Einstein action under this group of transformations is investigated. It is then shown that, if a modified volume element is included in the Hilbert-Einstein action, and the symmetry group extended to this modified volume element, the action is invariant with respect to the $\Omega$ transform. As a consequence of this symmetry it follows that all field theories of the free gravitation field derived from this superset are equivalent to the general theory of relativity. In other words, by a suitable choice of gauge the modified volume element may be eliminated. This choice would not be apparent without the inclusion of torsion in the connection. To illustrate the points made here the author relies heavily on the work of previous authors and retraces steps in previous work for clarity [1-3]. The work presented here is slightly modified version of work presented in the authors thesis [4].

## 2. The Ω transform

A general metric compatible connection, Γ on a space-time manifold, $M$, with metric, $g_{\mu\nu}$, may be expressed as,

$$\Gamma^{\alpha}{}_{\lambda\beta} = \left\{ {\alpha \atop \lambda\beta} \right\} + \frac{2}{N-1}\left(\delta^{\alpha}_{\lambda}T_{\beta} - g_{\lambda\beta}g^{\alpha\rho}T_{\rho}\right) + \tilde{K}^{\alpha}{}_{\lambda\beta} \qquad (4)$$

in which, $\left\{ {\alpha \atop \lambda\beta} \right\}$ is the Christoffel symbol, $T_{\beta}$ is the trace of the torsion tensor and $\tilde{K}^{\alpha}{}_{\lambda\beta}$ is the trace free part of the contortion tensor. Applying the transformation of Eq. (3) to the connection in Eq. (4) leads to a new connection,

$$\overline{\Gamma}^{\alpha}{}_{\lambda\beta} = \left\{ {\alpha \atop \lambda\beta} \right\} + \frac{2}{N-1}\left(\delta^{\alpha}_{\lambda}T_{\beta} - g_{\lambda\beta}g^{\alpha\rho}T_{\rho}\right) + \tilde{K}^{\alpha}{}_{\lambda\beta}$$
$$- \delta^{\alpha}_{\lambda}\partial_{\beta}\ln\Omega - \delta^{\alpha}_{\beta}\partial_{\lambda}\ln\Omega + g_{\lambda\beta}g^{\alpha\rho}\partial_{\rho}\ln\Omega \qquad (5)$$

A $\lambda$ transformation is applied to the connection in Eq. (5). By choosing $f = \ln\Omega$, we may eliminate the 5-th term in Eq. (5).

$$\overline{\overline{\Gamma}}^{\alpha}{}_{\lambda\beta} = \left\{ {\alpha \atop \lambda\beta} \right\} + \frac{2}{N-1}\left(\delta^{\alpha}_{\lambda}T_{\beta} - g_{\lambda\beta}g^{\alpha\rho}T_{\rho}\right) + \tilde{K}^{\alpha}{}_{\lambda\beta} - \delta^{\alpha}_{\lambda}\partial_{\beta}\ln\Omega + g_{\lambda\beta}g^{\alpha\rho}\partial_{\rho}\ln\Omega \qquad (6)$$

We may rewrite Equation (6) to give,

$$\overline{\overline{\Gamma}}^{\alpha}{}_{\lambda\beta} = \left\{ {\alpha \atop \lambda\beta} \right\} + \frac{2}{N-1}\left(\delta^{\alpha}_{\lambda}\left(T_{\beta} - \frac{N-1}{2}\partial_{\beta}\ln\Omega\right) - g_{\lambda\beta}g^{\alpha\rho}\left(T_{\rho} - \frac{N-1}{2}\partial_{\rho}\ln\Omega\right)\right) + \tilde{K}^{\alpha}{}_{\lambda\beta} \qquad (7)$$

Equation (7) suggests that we treat the trace of the torsion tensor as a gauge field. With this identification in place we have proven the following:

Given a space-time manifold, $M$ with $\dim(M) = N$ and $\text{sig}(M) = N-1$, with metric $g_{\mu\nu}$, and a metric compatible connection $\Gamma$, the transformation defined by,

$$g_{\mu\nu} \to \Omega^2 g_{\mu\nu} \tag{8}$$

$$T_\beta \to T_\beta + \frac{N-1}{2} \partial_\beta \ln \Omega \tag{9}$$

$$\Gamma^\alpha{}_{\mu\nu} \to \Gamma^\alpha{}_{\mu\nu} + \delta^\alpha_\nu \partial_\mu \ln \Omega \tag{10}$$

leaves the following quantities invariant.

$$\Gamma \xrightarrow{\Omega} \Gamma \tag{11}$$

$$R_{\mu\nu\alpha}{}^\beta \xrightarrow{\Omega} R_{\mu\nu\alpha}{}^\beta \tag{12}$$

$$R_{\mu\alpha} \xrightarrow{\Omega} R_{\mu\alpha} \tag{13}$$

The scalar curvature does not transform trivially under $\Omega$.

$$R = g^{\mu\nu} R_{\mu\nu} \xrightarrow{\Omega} \Omega^{-2} g^{\mu\nu} R_{\mu\nu} = \Omega^{-2} R \tag{14}$$

In addition to the invariance of $\{\Gamma, R_{\mu\nu\alpha}{}^\beta, R_{\mu\alpha}\}$ we may now show that the Einstein tensor is invariant.

$$R_{\mu\nu} - \frac{1}{2} g_{\mu\nu} R \xrightarrow{\Omega} R_{\mu\nu} - \frac{1}{2} \Omega^2 g_{\mu\nu} \Omega^{-2} R = R_{\mu\nu} - \frac{1}{2} g_{\mu\nu} R \tag{15}$$

It is worth mentioning that in the absence of the $\lambda$ transform the curvature tensors are invariant under the combination of a conformal transform and an SO(2) gauge transform applied to the torsion field.

**2. Hilbert-Einstein action**

The transformation presented in the previous section leaves the autoparallel structure of $M$ and the Riemann, Ricci and Einstein curvature tensors invariant. Despite this invariance, the scalar curvature as well as standard measures of length and volume is not invariant under this transformation.

For example, the Ω transformation produces a change in the standard linear action used to derive Einstein's free field equations. In the following $\gamma$ is a constant.

$$I = \gamma \int d^N x \sqrt{-g}\, R \tag{16}$$

$$I \xrightarrow{\Omega} \gamma \int d^4 x\, \Omega^{N-2} \sqrt{-g}\, R \tag{17}$$

The action is clearly invariant when $N = 2$. However it should be noted that the transformation applied to the action also includes terms which act on the torsion field contained in $R$.

Consider a class of volume elements in $N$-dimensions defined by,

$$dvol = h(x)\sqrt{-g}\, d^N x \tag{18}$$

in which $h(x)$ is a scalar under general coordinate transformations but considered here to obey the following transformation law under the Ω transformation defined in section 2.

$$h(x) \xrightarrow{\Omega} \Omega^S f(x) \tag{19}$$

Using this modified volume element we introduce a modified Hilbert-Einstein action,

$$I = \gamma \int d^N x\, h(x) \sqrt{-g}\, R \xrightarrow{\Omega} \gamma \int d^N x\, \Omega^{S+N-2} h(x) \sqrt{-g}\, R \tag{20}$$

The choice $S = 2 - N$ makes the modified action invariant under Ω. For convenience we choose $h(x) = e^{\kappa \Theta(x)}$, where $\kappa$ is a constant. With this definition in place, the Omega transformation is extended to include the scalar field $\Theta(x)$.

$$g_{\mu\nu} \xrightarrow{\Omega} \Omega^2 g_{\mu\nu} \tag{21}$$

$$T_\mu \xrightarrow{\Omega} T_\mu + \frac{N-1}{2} \partial_\mu \ln \Omega \tag{22}$$

$$\Theta \xrightarrow{\Omega} \Theta + \frac{2-N}{\kappa} \ln \Omega \tag{23}$$

The modified Hilbert-Einstein action and its derived field equations are invariant under this transformation.

## 3. Field equations and choice of gauge

In this section the field equations are derived from the modified Hilbert-Einstein action and the consequences of the internal symmetry presented in the last section are discussed.

The curvature tensors may be separated into separate metric induced and torsion induced parts.

$$R_{\mu\nu\alpha}{}^\beta = \mathfrak{R}_{\mu\nu\alpha}{}^\beta + D_\mu K^\beta{}_{\nu\alpha} - D_\nu K^\beta{}_{\mu\alpha} + K^\beta{}_{\mu\lambda} K^\lambda{}_{\nu\alpha} - K^\beta{}_{\nu\lambda} K^\lambda{}_{\mu\alpha} \tag{24}$$

$$R = \mathfrak{R} + 4 D^\alpha T_\alpha + \widetilde{K}^{\beta\alpha\lambda} \widetilde{K}_{\lambda\beta\alpha} + 4 \frac{N-2}{N-1} T_\lambda T^\lambda \tag{25}$$

in which $\mathfrak{R}$ is used to denote the part of the curvature tensor induced by the metric tensor only and the derivative $D$ is the metric compatible connection, containing only the Christoffel symbol.

The modified action now takes the form,

$$I = \gamma \int d^N x\, e^{\kappa\Theta} \sqrt{-g} \left( \mathfrak{R} + 4 D^\alpha T_\alpha + \widetilde{K}^{\beta\alpha\lambda} \widetilde{K}_{\lambda\beta\alpha} + 4 \frac{N-2}{N-1} T_\lambda T^\lambda \right) \tag{26}$$

Variation of equation (26) with respect to $\Theta$, $\widetilde{K}^{\beta\alpha\lambda}$, $T^\alpha$ and $g^{\mu\nu}$, leads to the following field equations respectively,

$$R = 0 \tag{27}$$

$$\widetilde{K}_{\beta\alpha\lambda} = 0 \tag{28}$$

$$\kappa \partial_\alpha \Theta = 2\frac{N-2}{N-1}T_\alpha \tag{29}$$

$$G_{\mu\nu} + \kappa\left(D_\mu \partial_\nu \Theta - g_{\mu\nu} D^2\Theta\right) + \kappa^2\left(\frac{-1}{N-2}\partial_\mu\Theta\partial_\nu\Theta + \frac{N-3}{2(N-2)}g_{\mu\nu}\partial_\sigma\Theta\partial^\sigma\Theta\right) = 0 \tag{30}$$

In the last field equation, equations (27) through (29) have been used to simplify terms. The trace of equation (30), and equation (27) together imply the following,

$$\mathfrak{R} = -2\kappa\frac{N-1}{N-2}D^2\Theta \tag{31}$$

$$\partial_\mu\Theta\partial^\mu\Theta = 0 \tag{32}$$

The scalar field in the modified volume element generates a Null-vector field in *M*, while the metric induced scalar curvature acts as a source for this scalar field. Finally note that when the original action is evaluated at the minimum it is identically zero.

The presence of the $\Omega$ symmetry in the action allows one to constrain a single element of the set of fields $\{\Theta, g_{\mu\nu}, T_\alpha\}$. Different choices of constraint will lead to different field equations, all of which are connected by a transformation. One choice is to fix the modified volume element, $\Theta$ = constant. This choice is consistent with the field equations and leads to the standard general theory of relativity, in which the Torsion field is identically zero.

## 4. Discussion and conclusion

By considering a more general connection than the Cristoffel symbols, and a more general action than the Hilbert-Einstein action, we have shown that an internal symmetry exists, which leaves the complete auto-parallel structure of the manifold and all field equations derived by a modified Hilbert-Einstein action unchanged. Additionally it has been shown that by a suitable choice of gauge all systems of this type are equivalent to those of the classical, general theory of relativity. If we had started with a metric field only, this symmetry would have been missed. Its presence was due to the immersion of the general relativity in a larger field theory, one which included a full torsion tensor and a modified volume element. Each of these modifications has been considered individually by authors in previous work. Torsion theory has a deep history; Einstein's lambda transformation, anti-symmetric second rank torsion potential and full third rank propagating torsion theory. Modified volume elements have been introduced to induce propagating torsion [3]. The primary result presented here is that when all these mechanisms are included together, that these theories are all equivalent to classical general relativity.

One may notice that the $\lambda$ transform was not necessary to imply the invariance of the field equation presented in section 3. Applying a conformal transform to the metric and an SO(2) gauge transform to the torsion trace are sufficient to show invariance of the modified Hilbert-Einstein action and the derived free field equations, However in this case the auto-parallel structure will no longer be invariant.

The effects of introducing matter fields, and higher derivative theory on this symmetry are currently under investigation.